\documentclass[journal]{IEEEtran}

\usepackage{lineno,hyperref}
\usepackage{bm}
\usepackage{mathrsfs}
\usepackage{amsmath}
\usepackage{graphicx}
\usepackage{epstopdf}
\usepackage{cite}
\usepackage{multirow}
\usepackage{amssymb}
\usepackage{booktabs}
\usepackage{xcolor}
\modulolinenumbers[5]

\ifCLASSINFOpdf

\else

\fi

\begin{document}

\title{Diffusion Multi-Rate LMS Algorithm for Acoustic Sensor Networks}

\author{Lu Lu,~\IEEEmembership{Member,~IEEE,}, Xiaomin Yang and Rongzhu Zhang

\thanks{Manuscript received March 29, 2020. This work was partially supported by the NSFC under Grant no. 61901285, 61901400, and 61701327, China Postdoctoral Science Foundation under Grant 2018M640916, Sichuan Science and Technology Fund under Grant 20YYJC3709, and Sichuan University Postdoctoral Interdisciplinary Fund.}
\thanks{L. Lu, X. Yang and R. Zhang are with the College of Electronics and Information Engineering, Sichuan University, Chengdu 610065, China. (e-mail: lulu19900303@126.com\;(L.\;Lu), arielyang@scu.edu.cn\;(X.\;Yang), zhang\_ rz@scu.edu.cn\;(R.\;Zhang).}}

\markboth{}
{Shell \MakeLowercase{\textit{et al.}}: Bare Demo of IEEEtran.cls for IEEE Journals}
\maketitle
\begin{abstract}
In this paper, we present a diffusion multi-rate least-mean-square (LMS) algorithm, named DMLMS, which is an effective solution for distributed estimation when two or more observation sequences are available with different sampling rates. Then, we focus on a more practical application in the wireless acoustic sensor networks (ASN). The filtered-x LMS (FxLMS) algorithm is extended to the distributed multi-rate system and it introduces collaboration between nodes following a diffusion strategy. Simulation results show that the effectiveness of the proposed algorithms.
\end{abstract}
%
\section{Introduction}
\label{sec:intro}

To overcome the limitations of the centralized strategy, the diffusion-based algorithms were developed for in-network distributed estimation \cite{sayed2014adaptive,cattivelli2010diffusion,lu2017diffusion,lu2018performance,zheng2017diffusion}. The diffusion least-mean-square (DLMS) algorithm is one of the most well-known distributed algorithms. By using the Adapt-then-Combine (ATC) policy, the DLMS algorithm allows measurement exchange and outperforms the global steepest-descent solution in some cases \cite{cattivelli2010diffusion}. The DLMS and its variant algorithms have been successfully applied to many applications, including position determination \cite{xia2016distributed}, active learning \cite{shen2016distributed}, decision-making \cite{tu2014distributed} and adaptive control \cite{yu2018prescribed}.

In some practical situations, when the adaptive algorithm is implemented, the input signal can be simultaneously obtained over many channels with different sample rates. For example, when the adaptive filter algorithm is used to speech signal processing, we may have a speech transmission available simultaneously over two digital channels: a high sample rate channel with low signal-to-noise ratio (SNR) and a low sample rate channel with high SNR \cite{therrien2002least,liu2017adaptive}. A similar problem is also encountered in image processing. We may have a suite of sensors to observe the infrared and visible-light simultaneously, but with differing resolutions \cite{therrien2002least}. Another example occurs in the adaptive filters based on subband techniques or a systolic array \cite{shynk1992frequency}. In this case, the algorithm also contains several input signals with different sample rates. In applications such as these, the used algorithm is a modified version of the least-mean-square (LMS) algorithm known as the multi-rate LMS (MLMS) algorithm in the literature \cite{shynk1992frequency,hawes2003lms,ukte2015adaptive}.

Multi-rate system is critical for amounts of image/signal processing applications. It is becoming increasingly important. The diffusion MLMS (DMLMS) algorithm proposed in this paper is excellent in solving multi-rate problem over distributed network, which can be used to achieve good estimation accuracy. The acoustic sensor networks (ASN) based on the distributed algorithm can be considered as a flexible and efficient solution for active noise control (ANC). The ANC problem over distributed network has been addressed in several previous studies, including \cite{antonanzas2015diffusion,ferrer2015active,plata2016incremental,song2016diffusion,kukde2017distributed}. In these works and other similar references on the topic, every node in ASN has single sample rate and less attention is paid to multi-rate distributed and in-network processing devices. In this work, the DMLMS algorithm is applied to the ASN problem, resulting the Filtered-x DMLMS (FxDMLMS) algorithm. Simulation results demonstrate that the DMLMS algorithm enjoys good performance for distributed estimation and the FxDMLMS algorithm has good noise reduction.

The paper is organized as follows. Section 2 describes distributed multi-rate system. In Section 3, the DMLMS algorithm is proposed. In Section 4, an application in ASN is presented to illustrate the effectiveness and advantage of the proposed FxMLMS algorithm. We show the computer simulation results in Section 5 and conclude the paper in Section 6.

\section{Problem formulation}
\label{sec:format}

\begin{figure}[!htb]
	\centering
	\includegraphics[scale=0.7] {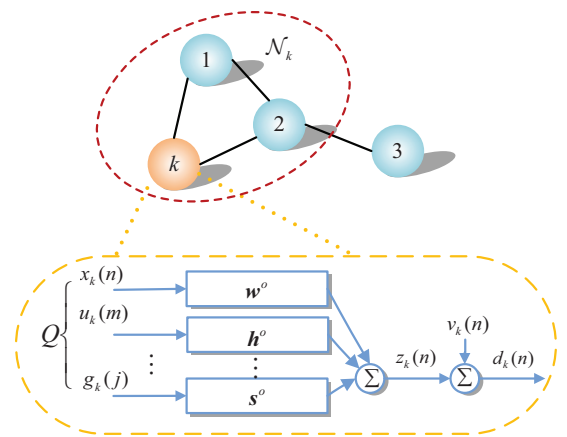}
	\caption{\label{1} Diagram of diffusion multi-rate system.}
	\label{Fig01}
\end{figure}

We consider the problem of identifying coefficients from a distributed network with $N$ sensor nodes indexed by $k \in \{1,\ldots, N\}$ (see Fig. \ref{Fig01}). For each node $k$, several different sampling rates are supposed to be input. Let $x_k(t)$ be the analog input at node $k$ and time $t$. Then, the analog signal $x_k(t)$ is sampled to several input signals: $x_k(n)$ with sample rate $f_{s_1}$, input $u_k(m)$ with sample rate $f_{s_2}$, and input $g_k(j)$ with sample rate $f_{s_Q}$. Define $m$, $n$ and $j$ are time indexes of low, moderate and high rate signals. At any time $n \in \mathbb{N}$, one can write
\begin{equation}
\begin{aligned}
n=&\;\Delta m+\delta,\\
j=&\;\Delta n+\delta,\;\;\delta=0,1,\ldots,\Delta-1,
\end{aligned}
\label{001}
\end{equation}
where $\mathbb{N}$ stands for the set of nature number. Thus, the input vector with different sample rates and length can be expressed as follows:
\begin{equation}
\bm x_k(n) = \left[x_k(n),x_k(n-1),\ldots,x_k(n-M_1+1)\right]^{\mathrm T}
\label{002}
\end{equation}
\begin{equation}
\bm u_k(m) = \left[u_k(m),u_k(m-1),\ldots,u_k(m-M_2+1)\right]^{\mathrm T}
\label{003}
\end{equation}
\begin{equation}
\bm g_k(j) = \left[g_k(j),g_k(j-1),\ldots,g_k(j-M_Q+1)\right]^{\mathrm T}
\label{004}
\end{equation}
where $M_1$, $M_2$ and $M_Q$ denote the length of the input vector. At every time instant $n$, node $k$ receives a new observation $\left\{{\bm x_k(n),\bm u_k(m),\ldots,\bm g_k(j),d_k(n)} \right\}$, where $z_k(n)$ is the system output, $d_k(n)$ is the scalar measurement (desired response), $\mathcal{N}_k$ is the set of nodes with which node $k$ shares information (including $k$ itself), and $v_k(n)$ is the additive noise. The output of the unknown multi-rate system at node $k$ can be expressed as
\begin{equation}
z_k(n) = \bm x^{\mathrm T}_k(n) \bm w^o + \bm u^{\mathrm T}_k(m) \bm h^o +,\ldots,+\bm g^{\mathrm T}_k(j)\bm s^o. 
\label{005}
\end{equation}
In this expression, $\bm w^o \in \mathbb{R}^{M_1 \times 1}$, $\bm h^o \in \mathbb{R}^{M_2 \times 1}$ and $\bm s^o \in \mathbb{R}^{M_Q \times 1}$ denote the coefficients of unknown system associated with the sample rate, where $\mathbb{R}$ denotes the set of real number. 

\section{DMLMS algorithm}
In this section, the DMLMS algorithm is proposed for coefficients estimation of the diffusion multi-rate system. Here, we only consider two sample rate in diffusion multi-rate system, that is, the input vector $\bm x_k(n)$ with low sample rate and the input vector $\bm u_k(m)$ with moderate sample rate. 

The error signal at node $k$ and time $n$ is defined as follow:
\begin{equation}
e_k(n) \triangleq d_k(n) - y_k(n)
\label{006}
\end{equation}
where $y_k(n)$ is the output of the DMLMS algorithm, which can be calculated by
\begin{equation}
y_k(n) = \bm x^{\mathrm T}_k(n) \hat{\bm w}_k(m) + \bm u^{\mathrm T}_k(m)\hat{\bm h}_k(m) 
\label{007}
\end{equation}
where $\hat {\bm w}_k(m)\in \mathbb{R}^{M_1 \times 1}$ and $\hat {\bm h}_k(m)\in \mathbb{R}^{M_2 \times 1}$ are the weight vector. By employing the ATC policy \cite{cattivelli2010diffusion}, the intermediate estimates $\hat {\bm \varphi}_k(m) \in \mathbb{R}^{M_1 \times 1}$ and $\hat {\bm \psi}_k(m) \in \mathbb{R}^{M_2 \times 1}$ are first updated as
\begin{equation}
\hat {\bm \varphi}_k(m+1) = \hat {\bm \varphi}_k(m) + \mu_1 e_k(n)\bm x_k(n) 
\label{008}
\end{equation}
\vspace{-5mm}
\begin{equation}
\hat {\bm \psi}_k(m+1) = \hat {\bm \psi}_k(m) + \mu_2 e_k(n)\bm u_k(m) 
\label{009}
\end{equation}
where $\mu_1$ and $\mu_2$ are the step size (learning rate). Replacing the intermediate estimate in (\ref{008}) and (\ref{009}) by the linear-combination estimate, we have
\begin{equation}
\hat {\bm \varphi}_k(m+1) = \hat {\bm w}_k(m) + \mu_1 e_k(n)\bm x_k(n)
\label{010}
\end{equation}
\vspace{-5mm}
\begin{equation}
\hat {\bm \psi}_k(m+1) = \hat {\bm h}_k(m) + \mu_2 e_k(n)\bm u_k(m).  
\label{011}
\end{equation}
Such substitutions are reasonable since at a given time $m$, the weight parameters $\hat {\bm w}_k(m)$ and $\hat {\bm h}_k(m)$, contain more information than the corresponding intermediate estimates $\hat{\bm \varphi}_k(m)$ and $\hat{\bm \psi}_k(m)$, and therefore can enhance the estimation performance of the proposed algorithm. Furthermore, these substitutions avoid the initialization of the new parameters \cite{cattivelli2010diffusion}.

Then, the final weight vector and feedback weight are estimated through the linear combinations as follows:
\begin{equation}
\hat{\bm w}_k(m+1) = \sum\limits_{l \in {\mathcal{N}_k}} {c_{l,k} \hat{\bm \varphi}_l(m+1)},
\label{012}
\end{equation}
\vspace{-2mm}
\begin{equation}
\hat{\bm h}_k(m+1) = \sum\limits_{l \in {\mathcal{N}_k}} {c_{l,k} \hat{\bm \psi}_l(m+1)},
\label{013}
\end{equation}
where $c_{l,k} \ge 0$ are the non-negative real constants, and satisfy:
\begin{equation}
c_{l,k} = 0\;\;\mathrm {if}\;l \notin {\mathcal{N}_{k,\;\;}}\mathrm {and}\;\;\sum\limits_{k=1}^N {c_{l,k} = 1.}
\label{014}
\end{equation}
Compared to its centralized counterparts, the diffusion strategy does not need to utilzie any network process among sensors. Moreover, it does not require a central processor for calculating, which reduces the computational complexity in distributed networks. In general, the diffusion strategy is suitable for the multi-rate distributed system.

\section{FxDMLMS algorithm}

\begin{figure}[!htb]
	\centering
	\includegraphics[scale=0.6] {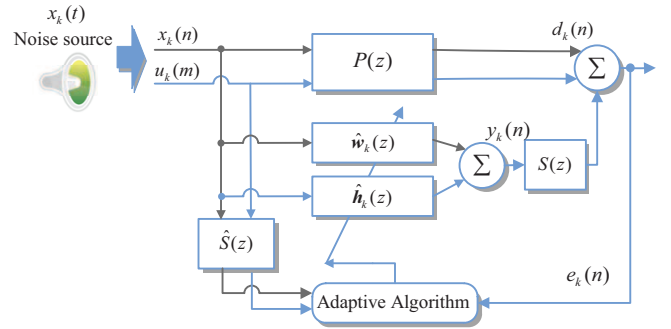}
	\caption{\label{2} The block diagram of FxMLMS for ANC system at node $k$.}
	\label{Fig02}
\end{figure}

In this section, the DMLMS algorithm is applied to ASN for active noise control. The block diagram of ANC system with multi-rate is illustrated in Fig. \ref{Fig02}, where $x_k(t)$ represents the analog noise source, $x_k(n)$ denotes the noise signal with low sample rate, $u_k(m)$ denotes the noise signal with high sample rate, $P(z)$ denotes the primary path from the noise signals (reference microphone signals) $x_k(n)$ and $u_k(m)$ to the error microphone $e_k(n)$, $S(z)$ is the secondary path transfer function between the output of the adaptive controllers $\hat{\bm w}_k(z)$ and $\hat{\bm h}_k(z)$ and the output of the error microphone $e_k(n)$, and $y_k(n)$ is the output of the controller. In this paper, the secondary path $S(z)$ and its estimate $\hat S(z)$ are assumed to be the same.

The error signal at node $k$ and time $n$ can be defined as
\begin{equation}
e_k(n) \triangleq d_k(n) - s(n)*y_k(n)
\label{015}
\end{equation}
where $s(n)$ is the impulse response of the secondary path transfer function $S(z)$, and $*$ is the linear convolution operator. The controller's output is calculated by
\begin{equation}
y_k(n) = \bm x^{\mathrm T}_k(n)\hat{\bm w}_k(m) + \bm u^{\mathrm T}_k(m)\hat{\bm h}_k(m).
\label{016}
\end{equation}
The definitions of the $\bm x_k(n)$, $\bm u_k(m)$, $\hat{\bm w}_k(m)$ and $\hat{\bm h}_k(m)$ can be found in section 3. By extending DMLMS algorithm in ASN, the adaptation step of the FxDMLMS algorithm can be expressed as
\begin{equation}
\hat{\bm \varphi}_k(m+1) = \hat{\bm \varphi}_k(m) + \mu_1 e_k(n)\bm {x'}_k(n) 
\label{017}
\end{equation}
\vspace{-5mm}
\begin{equation}
\hat{\bm \psi}_k(m+1) = \hat{\bm \psi}_k(m) + \mu_2 e_k(n)\bm {u'}_k(m) 
\label{018}
\end{equation}
where
\begin{equation}
\bm {x'}_k(n) = \bm {x}_k(n)*s(n), 
\label{019}
\end{equation}
\vspace{-5mm}
\begin{equation}
\bm {u'}_k(m) = \bm {u}_k(m)*s(n). 
\label{020}
\end{equation}
Similarly, we can replace the intermediate estimate by the linear combination estimation in (\ref{012}) and (\ref{013})
\begin{equation}
\hat{\bm \varphi}_k(m+1) = \hat{\bm w}_k(m) + \mu_1 e_k(n)\bm {x'}_k(n) 
\label{021}
\end{equation}
\vspace{-5mm}
\begin{equation}
\hat{\bm \psi}_k(m+1) = \hat{\bm h}_k(m) + \mu_2 e_k(n)\bm {u'}_k(m). 
\label{022}
\end{equation}
The communication with neighbor nodes in ASN is valuable for suppression noise. The $k$th node employs the information of the $(k -1)$th node in the previous iteration to calculate its own estimation in the current iteration. Therefore, the information of all the nodes spreads out over the ASN at each iteration. 

\section{Simulation results}

We present simulation results to verify the effectiveness of the proposed algorithms in the context of distributed estimation and active noise control in ASN. For distributed estimation problem, the performance of algorithms are evaluated in terms of the averaged-network excess mean-square error (EMSE)
\begin{equation}
\mathrm{network\;EMSE} = \frac{1}{N}\sum\limits^N_{k=1} (z_k(n)-y_k(n))^2. 
\label{023}
\end{equation}
The performance of the algorithms in ASN is quantified by noise reduction (NR)
\begin{equation}
\mathrm{NR} = \frac{1}{N}\sum\limits^N_{k=1} (d_k(n)-y_k(n))^2. 
\label{024}
\end{equation}
All simulations are averaged over 100 independent trials. 

\begin{figure}[!htb]
	\centering
	\includegraphics[scale=0.45] {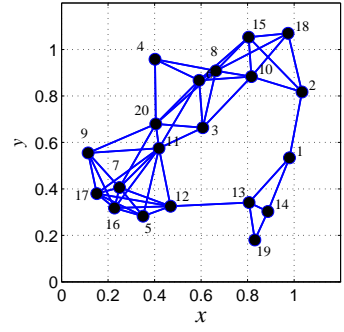}
	\caption{\label{3} Network topology.}
	\label{Fig03}
\end{figure}

\begin{figure}[!htb]
	\centering
	\includegraphics[scale=0.48] {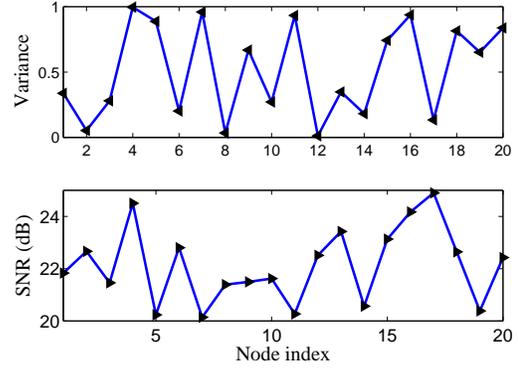}
	\caption{\label{4} The variance of the input signal and SNR at all nodes in the nonlinear distributed network.}
	\label{Fig04}
\end{figure}

\begin{figure}[!htb]
	\centering
	\includegraphics[scale=0.6] {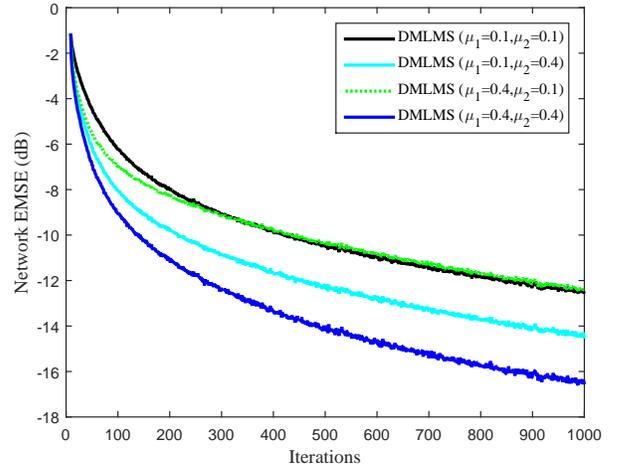}
	\caption{\label{5} Network EMSE of the DMLMS algorithm versus different step sizes.}
	\label{Fig05}
\end{figure}

\begin{figure}[!htb]
	\centering
	\includegraphics[scale=0.6] {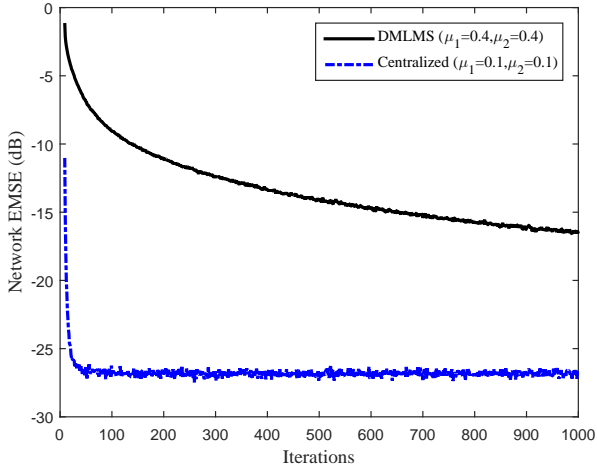}
	\caption{\label{6} Network EMSE learning curve.}
	\label{Fig06}
\end{figure}

\begin{figure}[!htb]
	\centering
	\includegraphics[scale=0.45] {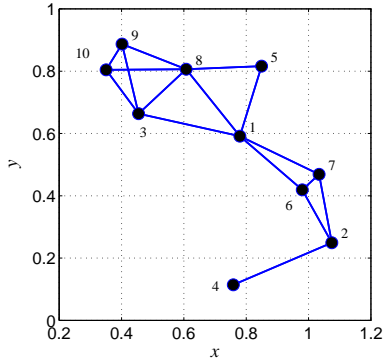}
	\caption{\label{7} Network topology of ASN.}
	\label{Fig07}
\end{figure}

\begin{figure}[!htb]
	\centering
	\includegraphics[scale=0.45] {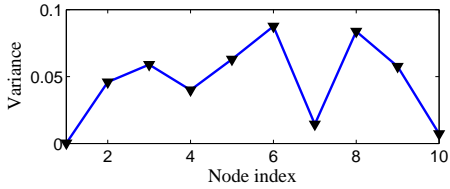}
	\caption{\label{8} Variance of the noise source over the node.}
	\label{Fig08}
\end{figure}

\begin{figure}[!htb]
	\centering
	\includegraphics[scale=0.5] {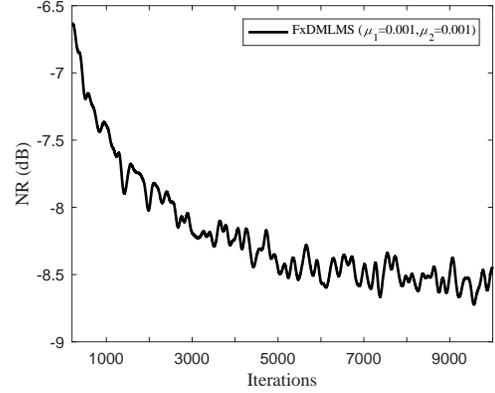}
	\caption{\label{9} NR learning curve for ANC system.}
	\label{Fig09}
\end{figure}

\subsection{Distributed estimation}
In the first example, the distributed network is composed of 20 nodes, as shown in Fig. \ref{Fig03}. The linear combination coefficients $\{{c_{l,k}}\}$ in (\ref{012})-(\ref{013}) are selected based on the metropolis rule \cite{cattivelli2010diffusion}. The variances of the input $x_k(t)$ and SNR are shown in Fig. \ref{Fig04}. We set $M_1=5$, $M_2=4$, $\Delta=2$ and $\delta=0$. First, we investigate the effect of the step size on the performance of the algorithm. Fig. \ref{Fig05} shows the network EMSE of the DMLMS algorithm versus different $\mu_1$ and $\mu_2$. As can be seen, the DMLMS algorithm achieves the best performance with $\mu_1=0.4$ and $\mu_2=0.4$. Fig. \ref{Fig06} compares the DMLMS algorithm and its centralized version \footnote{A centralized (global) version is not distributed in nature, it requires acces to data $\left\{{\bm x_k(n),\bm u_k(m),\ldots,\bm g_k(j),d_k(n)} \right\}$ across the entire network. Hence, it requires a fusion center in network, which prohibits its practical applications.} for distributed estimation. As it is shown, the centralized algorithm has the faster convergence rate and smaller misadjustment than the DMLMS algorithm. However, the performance of the centralized network is adversely more severely than in a distributed set-up when applied to large networks. 

\subsection{Active noise control in ASN}

In the second example, the length of weight vector of $\hat{\bm w}_k(n)$ is set as 16, the length of weight vector of $\hat{\bm h}_k(n)$ is set as 8, $\delta=0$, $\Delta=2$, the primary path $P(z)$ and the secondary path $S(z)$ are modeled as FIR filter of length 16 and 8, which is generated randomly \cite{song2016diffusion}. The ASN is composed of 10 nodes, as shown in Fig. \ref{Fig07}. A Gaussian noise with zero mean is employed as the primary input source, and the variance of the noise source is shown in Fig. \ref{Fig08}. Fig. \ref{Fig09} shows the network EMSE learning curve for ANC system. As can be seen, the proposed FxDMLMS algorithm achieves good performance for ASN. Note that the distributed algorithms in \cite{ferrer2015active,plata2016incremental} are the incremental version, which are characterized by a heavy computation burden. The algorithm in \cite{song2016diffusion} is based on the diffusion strategy. All the existing algorithms have different structures as compared with the FxDMLMS algorithm. For these reasons, we decide only show the learning curve of the proposed FxDMLMS algorithm in Fig. \ref{Fig09}.

\section{Conclusion}

In this paper, we proposed a diffusion multi-rate LMS algorithm named DMLMS, which was derived by extending the MLMS algorithm to distributed network. One of the advantages of the DMLMS algorithm is that it has only two parameters (step sizes), which make the algorithm easy to implementation. Based on the concept of the DMLMS algorithm, we further proposed the DMFxLMS for active noise control problem in ASN. Simulation results showed that the effective performance of the DMLMS and DMFxLMS algorithms. In the future study, we will focus on solving the impulsive noise source in ASN. The cost function proposed in \cite{lu2018ann} will be cosidered for use.



\bibliographystyle{IEEEtran}
\bibliography{IEEEabrv,mybibfile}

\end{document}